\documentclass[aps,prb,twocolumn,showpacs,psfig,superscriptaddress,longbibliography]{revtex4-1}

\usepackage{textcomp}
\usepackage{times}
\usepackage{graphicx}
\usepackage{float}
\usepackage{latexsym,amsmath,amssymb,bm,euscript}
\usepackage{color}
\usepackage{subfigure}
\usepackage{epstopdf}
\usepackage[colorlinks=true,linkcolor=blue,citecolor=blue]{hyperref}
\usepackage{hyperref}
\usepackage{soul}
\usepackage[normalem]{ulem}
\usepackage{mathrsfs}
\usepackage{amsmath}
\usepackage{lettrine}
\usepackage{xspace}
\usepackage{textcomp}

\begin{document}

\hyphenpenalty=5000
\tolerance=1000

\title{Field-induced magnetic quantum phase transitions in a bilayer triangular lattice antiferromagnet K$_2$Co$_2$(HPO$_3$)$_2$(C$_2$O$_4$)(H$_2$O)$_2$}	

\author{Yiqiu Liu}
\thanks{These authors contributed equally to this study.}
\affiliation{School of Mathematics and Physics,~North China Electric Power University,~Beijing,~102206,~China}

\author{Chengcheng Yin}
\thanks{These authors contributed equally to this study.}
\affiliation{School of Chemistry and Chemical Engineering $\&$ Wuhan High Magnetic Field Center, Huazhong University of Science and Technology, Wuhan 430074, China}

\author{Xinyu Shi}
\thanks{These authors contributed equally to this study.}
\affiliation{School of Physi and Beijing Key Laboratory of Opto-electronic Functional Materials $\&$ Micro-nano Devices, Renmin University of China, Beijing, 100872, China}

\author{Xiaohui Bo}
\thanks{These authors contributed equally to this study.}
\affiliation{School of Physics and Beijing Key Laboratory of Opto-electronic Functional Materials $\&$ Micro-nano Devices, Renmin University of China, Beijing, 100872, China}

\author{Kefan Du}
\affiliation{School of Physics and Beijing Key Laboratory of Opto-electronic Functional Materials $\&$ Micro-nano Devices, Renmin University of China, Beijing, 100872, China}

\author{Zhanlong Wu}
\affiliation{School of Physics and Beijing Key Laboratory of Opto-electronic Functional Materials $\&$ Micro-nano Devices, Renmin University of China, Beijing, 100872, China}

\author{Rui Bian}
\affiliation{School of Physics and Beijing Key Laboratory of Opto-electronic Functional Materials $\&$ Micro-nano Devices, Renmin University of China, Beijing, 100872, China}

\author{Shuo Li}
\affiliation{Institute of Physics, Chinese Academy of Sciences, and Beijing National Laboratory for Condensed Matter Physics, Beijing 100190, China}  

\author{Jun Luo}
\affiliation{Institute of Physics, Chinese Academy of Sciences, and Beijing National Laboratory for Condensed Matter Physics, Beijing 100190, China}  

\author{Jie Yang}
\affiliation{Institute of Physics, Chinese Academy of Sciences, and Beijing National Laboratory for Condensed Matter Physics, Beijing 100190, China}

\author{Yanhong Wang}
\affiliation{School of Chemistry and Chemical Engineering $\&$ Wuhan High Magnetic Field Center, Huazhong University of Science and Technology, Wuhan 430074, China} 

\author{Weiqiang Yu}
\email{wqyu\_phy@ruc.edu.cn}
\affiliation{School of Physics and Beijing Key Laboratory of Opto-electronic Functional Materials $\&$ Micro-nano Devices, Renmin University of China, Beijing, 100872, China}
\affiliation{Key Laboratory of Quantum State Construction and Manipulation (Ministry of Education), Renmin University of China, Beijing, 100872, China}

\author{Rui Zhou}
\email{rzhou@iphy.ac.cn}
\affiliation{Institute of Physics, Chinese Academy of Sciences, and Beijing National Laboratory for Condensed Matter Physics, Beijing 100190, China}

\author{Jinshan Zhang}
\email{zhangjs@ncepu.edu.cn}
\affiliation{School of Mathematics and Physics,~North China Electric Power University,~Beijing,~102206,~China}

\author{Yi Cui}
\email{cuiyi@ruc.edu.cn}
\affiliation{School of Physics and Beijing Key Laboratory of Opto-electronic Functional Materials $\&$ Micro-nano Devices, Renmin University of China, Beijing, 100872, China}
\affiliation{Key Laboratory of Quantum State Construction and Manipulation (Ministry of Education), Renmin University of China, Beijing, 100872, China}

\author{Hongcheng Lu}
\email{hcl@hust.edu.cn}
\affiliation{School of Chemistry and Chemical Engineering $\&$ Wuhan High Magnetic Field Center, Huazhong University of Science and Technology, Wuhan 430074, China}

\begin{abstract}

We present a $^{31}$P nuclear magnetic resonance (NMR) study on the spin-$1/2$ bilayer triangular-lattice antiferromagnet K$_2$Co$_2$(HPO$_3$)$_2$(C$_2$O$_4$)(H$_2$O)$_2$. Measurements were performed at temperatures down to 0.06~K and magnetic fields up to 26~T. In the field range of 0.5~T to 4~T, the low-temperature spin-lattice relaxation rate $1/T_1$ exhibits a peak, signaling the onset of a magnetically ordered state. Within this ordered phase, $1/T_1$ shows gapless behavior, suggesting a spin supersolid phase. This ordering is suppressed around 4~T. Between 6~T and 18~T, clear splittings of the NMR spectra are observed, indicating the emergence of a second ordered magnetic phase. The field derivative of the low-temperature Knight shift $K_{\rm n}$ reveals changes in slope at approximately 10~T, 15.5~T, and 18~T. The first two features correspond to bulk magnetizations of $1/3$ and $2/3$ of the fully saturated moment, respectively, while the last one marks a quantum phase transition from the second ordered phase to the fully polarized state. These results establish the studied compound as a platform for investigating field-induced quantum phase transitions and emergent phases in frustrated bilayer triangular-lattice antiferromagnets.



\end{abstract}

\maketitle

\section{Introduction}

Triangular Lattice Antiferromagnets (TLAFMs), as a typical platform for geometric frustration, are ideal for realizing quantum spin liquid (QSL)~\cite{Wietek_PRX_2024,Diep_CRP_2025,Li_2025_CPL}. The classical Heisenberg model predicts a $120^{\circ}$ long-range magnetic order at zero temperature for an isotropic nearest-neighbor Hamiltonian~\cite{Kawamura_JAP_1987,Chubukov_PRB_1992}, but quantum fluctuations (especially for the spin $S=1/2$ system) may stabilize the QSL state~\cite{Fortune_PRL_2009,Xie_NPJ_2023,Park_NC_2021,Willsher_PRR_2025}. 
However, the ground state of real materials is highly sensitive to additional interactions beyond the ideal model. For instance, the introduction of spatial anisotropy, further-neighbor exchanges such as next-nearest-neighbor coupling $J_{2}$, or bond anisotropy and spin dimerization~\cite{Muller_NC_2024} can significantly modify the phase diagram, leading to a variety of competing phases such as stripe order~\cite{Maksimov_PRX_2019,Li_PRB_2020}, nematic states~\cite{Park_PRL_2008,Starykh_RPP_2015}, and QSLs~\cite{li_SR_2015,Li_PRL_2016,Muller_NC_2024}. This makes finding a pure $S=1/2$ triangular lattice QSL exceptionally challenging, as exemplified by the significant structural disorder in YbMgGaO$_4$~\cite{li_SR_2015,Lima_JPCM_2020,Rao_NC_2021,Shen_NC_2018,Zhang_2024_CPL}. 

Recently, cobalt-based triangular lattice materials, such as BaCo$_{2}$(AsO$_{4}$)$_{2}$~\cite{Maksimov_PRB_2022,Halloran_PNSUA_2023,Zhang_NM_2023,Bose_PRB_2023,tu_CPL_2025} and Na$_{2}$BaCo(PO$_{4}$)$_{2}$~\cite{Sheng_PNAS_2022,Li_NC_2020,Hussain_PRB_2025},  have emerged as a new frontier in the study of quantum magnetism. In particular, the recently synthesized bilayer compounds such as K$_2$Co$_2$(SeO$_3$)$_3$~\cite{Effenberger_ZK_1999} and Rb$_2$Co$_2$(SeO$_3$)$_3$ present distinct systems with Co$^{2+}$ ions forming face-sharing dimers on the triangular lattice~\cite{Xu_CM_2024}. This intrinsic dimerization introduces a new scale of bond-dependent interaction, adding further complexity to the already rich interplay of anisotropy and frustration. In this system, strong spin-orbit coupling and crystal field effects lead to strong Ising-like dimerized magnetic couplings~\cite{Winter_JFM_2022,Liu_PRB_2023}. 
Such systems provide a fertile ground for unconventional quantum phases, including Berezinskii–Kosterlitz–Thouless (BKT) phases at zero field~\cite{ChenGang_PRL_2024} and field-induced magnetization plateaus whose microscopic nature remains under active investigation~\cite{Yong_SSC_2014,Honecker_JP_1999,Plumer_PA_1995}.

In this work, we present a nuclear magnetic resonance (NMR) study on a new bilayer triangular-lattice antiferromagnet with effective spin-$1/2$, K$_2$Co$_2$(HPO$_3$)$_2$(C$_2$O$_4$)(H$_2$O)$_2$. In this compound, the bilayers are well separated by organic ligands, which should significantly reduce interlayer coupling and thus suppress three-dimensional magnetic ordering. This makes the system a promising platform for exploring novel quantum phases in bilayer triangular-lattice antiferromagnets. We performed NMR measurements at temperatures down to 0.06~K and magnetic fields up to 26~T. By using NMR line splittings to probe magnetic ordering and the spin-lattice relaxation rate to characterize low-energy spin fluctuations, we find that the material exhibits a series of ordered phases in magnetic fields up to 18~T. In particular, the low-field ordered phase exhibits gapless spin excitations, whereas the high-field ordered phase is gapped. Furthermore, the magnetization inferred from the Knight shift reveals plateaus at approximately $1/3$ and $2/3$ of the saturated moment, which are consistent with the gapped excitations observed in the corresponding field regions. The $2/3$ plateau is rarely observed in single-layer TLAFMs. Our results demonstrate that the bilayer system provides a new platform for investigating exotic quantum phases and field-induced quantum phase transitions.

\section{Materials and Techniques}

\begin{figure}[t]
\includegraphics[width=8.5cm]{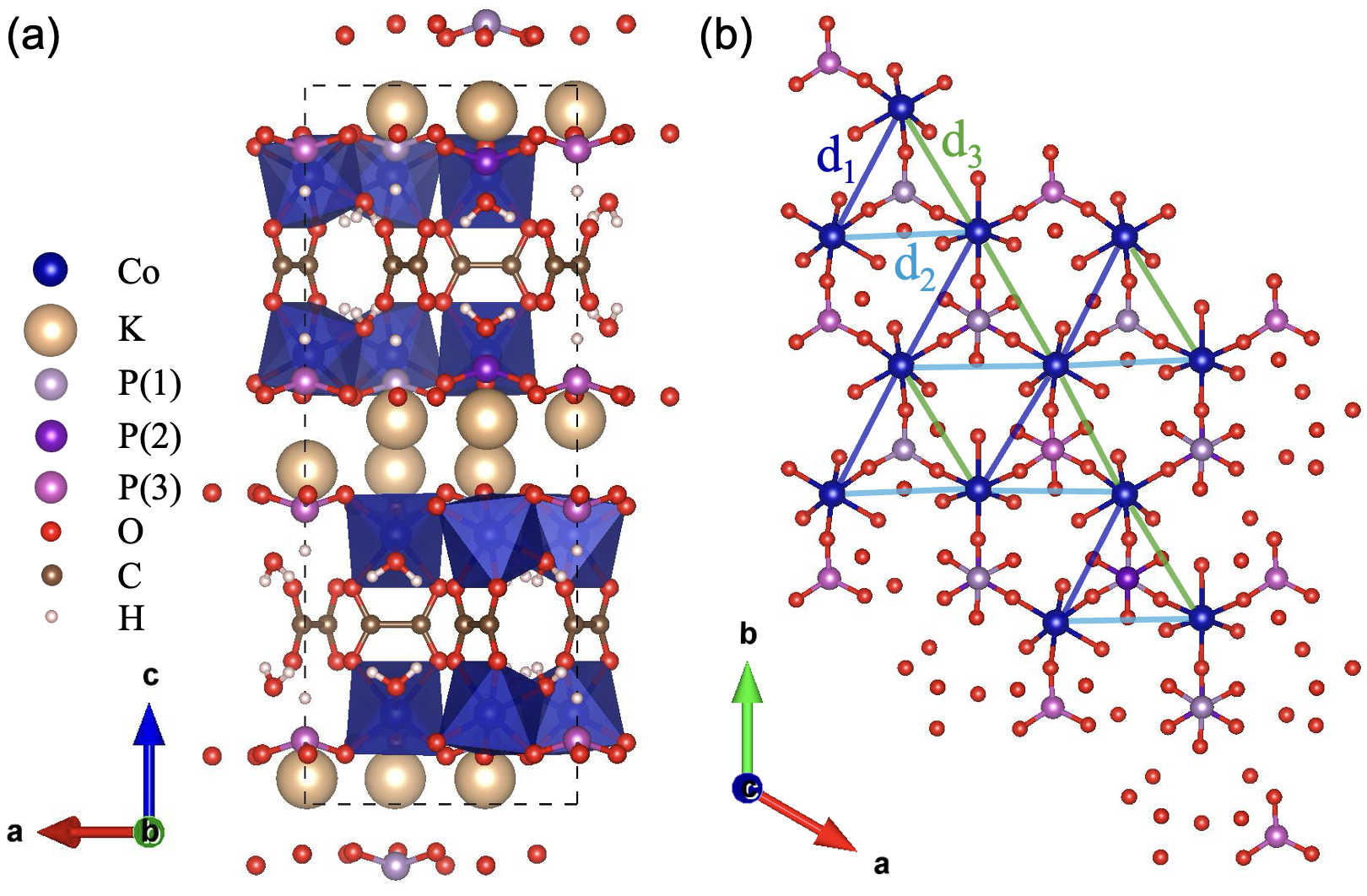}
\caption{\label{struc}~\textbf{Crystal structure of K$_2$Co$_2$(HPO$_3$)$_2$(C$_2$O$_4$)(H$_2$O)$_2$.}
(a) A schematic drawing of the 3D crystal structure with an AA stacking of the triangular lattice within the bilayer, and an AB stacking among the bilayers.
(b) Top view of a single triangular layer of Co$^{2+}$ ions. The lattice is slightly distorted, with inequivalent Co$^{2+}$-- Co$^{2+}$ bond lengths $d_1$, $d_2$, and $d_3$, indicating a non-equilateral triangular lattice.
}
\end{figure}

The polycrystalline sample K$_2$Co$_2$(HPO$_3$)$_2$(C$_2$O$_4$)(H$_2$O)$_2$ was synthesized by the hydrothermal method~\cite{Lu_2026}. As shown in Fig.~\ref{struc}(a), the compound crystallizes in a bilayer structure with an AB stacking sequence, and K$^{+}$ ions are located between the bilayers. Each bilayer consists of two planes of CoO$_6$ octahedra in a triangular lattice aligned along the $c$ axis. Therefore, the material can be treated as an $S=1/2$ bilayered triangular lattice magnet formed by Co$^{2+}$ dimers. A top view of one triangular layer within the bilayer is shown in Fig.~\ref{struc}(b). Note that the triangular lattice is slightly distorted, with inequivalent Co$^{2+}$-Co$^{2+}$ distances of $d_1$ = 5.448~\AA, $d_2$ = 5.430~\AA, and $d_3$ = 5.755~\AA~\cite{Lu_2026}. 
Three inequivalent $^{31}$P sites, denoted as P(1), P(2), and P(3), are located in the interstitial regions between bilayers.

The NMR measurements were performed on $^{31}$P nuclei, which have a spin $I=1/2$ and a Zeeman factor $\gamma$~$=$~17.235~MHz/T. A 12~Tesla magnet was used to perform measurements with temperature down to 0.06~K in a dilution refrigerator. The high field measurements were performed in an all-superconducting magnet at the Synergetic Extreme Condition User Facility (SECUF) with field up to 26~T and temperature down to 1.8~K. The magnetic field was calibrated by the resonance frequency of $^{27}$Al foil. 

The NMR spectra were obtained by the standard spin-echo technique, with a typical $\pi/2$ pulse length of 2$\mu$s. For broad spectra at low temperatures, the whole spectra were acquired by sweeping the measurement frequency. The Knight shift $K_{\rm n}$ was calculated by $K_{\rm n}$~$=$~$(f/{\gamma}H-1){\times}100\%$, where $f$ is the resonance frequency.
The spin-lattice relaxation rate $1/T_1$ was measured using the spin inversion-recovery method, with the nuclear magnetization fitted to the exponential function: $M(t)$~$=~$$M(\infty)[1-be^{-(t/T_1)^\beta}]$ with $1<b<2$, where $\beta$ is a stretching factor. In the paramagnetic (PM) phase, $\beta$~$\approx$~1, indicating high quality of the material. Note that with $1/T_1$~$=~$$T\sum_q~A_{hf}^{2}(q){\text{Im}~\chi(q,\omega{\rightarrow}0)}/{\omega}$, $1/T_1$ probes low-energy spin dynamics. Here $A_{hf}(q)$ is the hyperfine coupling structure factor, $\chi$ is the dynamic susceptibility of the electrons, $q$ is the wave vector, and $\omega$ is the NMR frequency.

\section{Low-field NMR measurements}

\subsection{NMR spectra and Knight shift}


\begin{figure}[t]
\includegraphics[width=8.5cm]{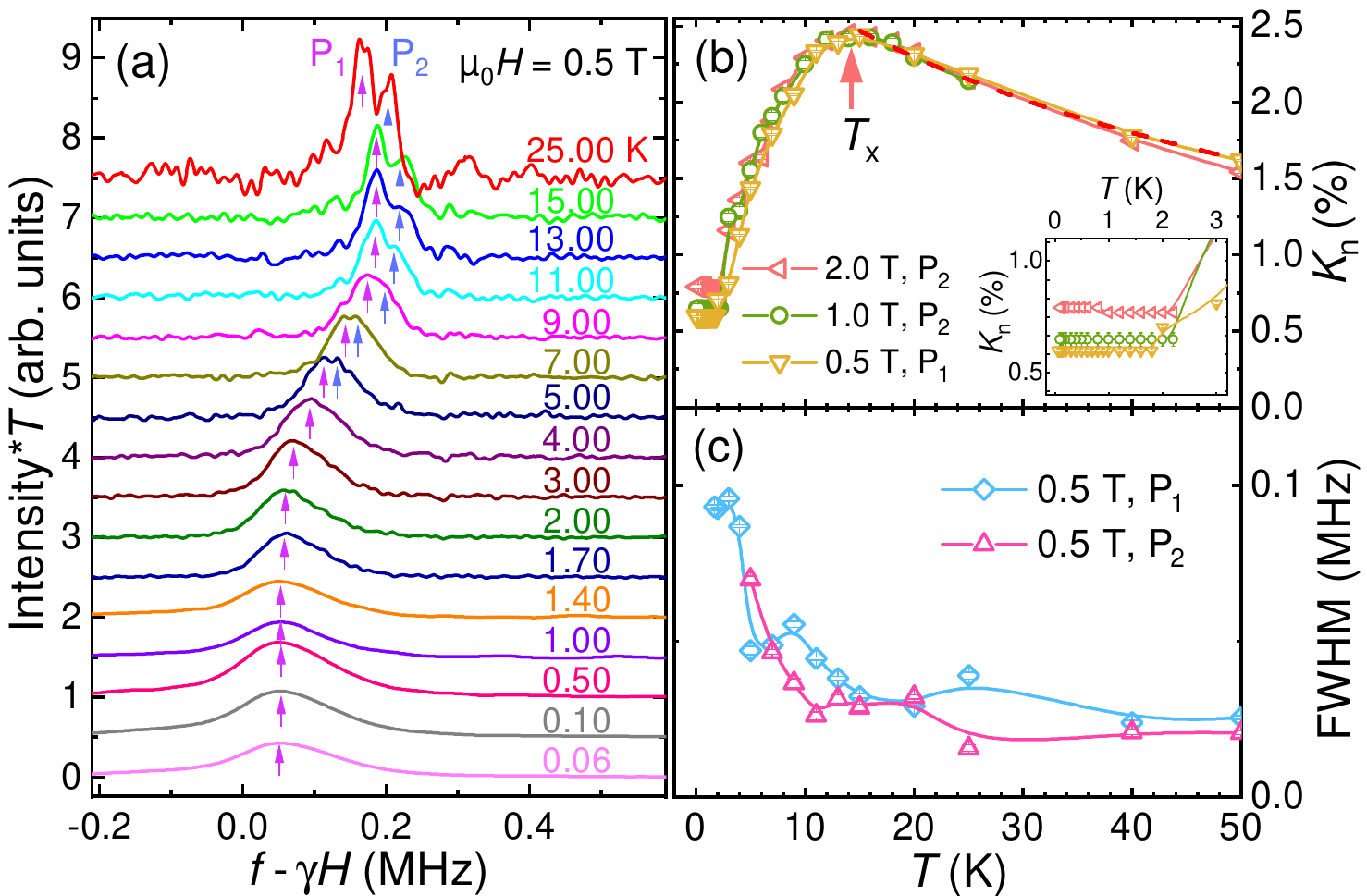}
\caption{\label{spec}~\textbf{NMR spectra and Knight shift with field from 0.5~T to 2~T.}
(a) Spectra at a fixed field of 0.5~T measured at selected temperatures. Vertical offsets are applied for clarity. Up arrows mark the peak positions, labeled as $\rm{P}_1$ and $\rm{P}_2$,  in the spectra.  
(b) Knight shift $K_{\rm n}$ as functions of temperatures measured at different fields, determined on $\rm{P}_1$ or $\rm{P}_2$.  The red dotted line is the Curie-Weiss law fit to $K_{\rm n}$ in the temperature range of 15~K to 50~K. The up arrow marks the peak position in $K_{\rm n}$ at a temperature labeled $T_{x}$. Inset: Enlarged view of the low-temperature data.  
(c) FWHM of the spectra as functions of temperatures, obtained by the double-Lorentz fit at temperatures above 5~K. Below 5~K, a single-Lorentz fit is applied instead, as $\rm{P}_1$ and $\rm{P}_2$ overlap and are not differentiable by the double-Lorentz fit.
}
\end{figure}

To further investigate the low-field phase, we focus on the temperature evolution of the spectra at 0.5~T, as shown in Fig.~\ref{spec}(a). 
At temperatures above 5~K, a double-peak feature, labeled as $\rm{P}_1$ and $\rm{P}_2$, is resolved in the spectra. The relative spectral weight of $\rm{P}_1$ and $\rm{P}_2$ is approximately 2:1, consistent with the presence of three inequivalent $^{31}$P sites in the crystal structure. 
Upon cooling from 25~K, the peak frequencies of $\rm{P}_1$ and $\rm{P}_2$ first move toward higher frequency, and then toward lower frequency at temperatures below 13~K. The NMR linewidth also increases with decreasing temperature, leading to a single peak due to the overlap of $\rm{P}_1$ and $\rm{P}_2$ below 5~K. 

We then applied a double-Lorentz fit and a single-Lorentz fit to the spectra at temperatures above and below 5~K respectively to resolve the resonance frequency and the linewidth of each peak. The Knight shift $K_{\rm n}$ is then calculated from the peak frequency on $\rm{P}_1$ or $\rm{P}_2$ at low fields, and then plotted as functions of temperatures in Fig.~\ref{spec}(b). 
$K_{\rm n}(T)$ can also be fitted to the Curie-Weiss form $K_{\rm n}(T)\sim1/(T+{\theta_{\rm CW}})$ at high temperatures with Curie-Weiss temperature $\theta_{\rm CW}~{\approx}$~52.3 K. 

Upon cooling, $K_{\rm n}$ first increases and then decreases, exhibiting a peaked feature at about 15~K. 
The downturn of the uniform susceptibility is usually caused by the onset of dimerization, AFM ordering, or a quantum disordered phase such as a spin liquid. In order to determine this, the full-width-at-half-maximum (FWHM) at each temperature is also obtained from the line fits and plotted in Fig.~\ref{spec}(c). The FWHM stays nearly constant above 20~K and then increases below 15~K, which is the same temperature at which $K_{\rm n}$ peaks. Such an increase of the FWHM in principle is consistent with either a symmetry breaking or a short-range order, rather than dimerization or quantum spin liquid. We will discuss this further with the spin-lattice relaxation data below.


\subsection{Spin-lattice relaxation rate}

\begin{figure}[t]
\includegraphics[width=7.5cm]{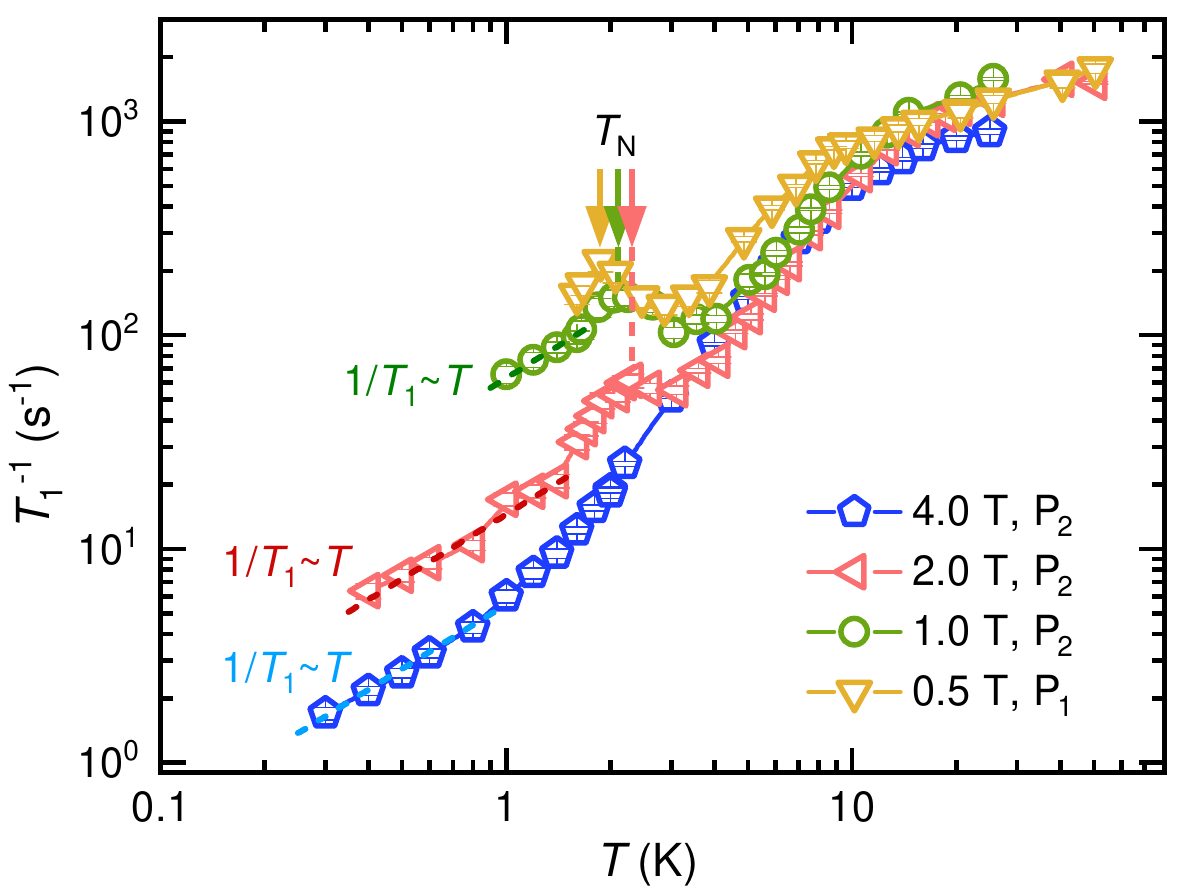}
\caption{\label{slrr}~\textbf{Spin-lattice relaxation rate with field up to 2~T.} 
$1/T_1$ measured on $\rm{P}_1$ or $\rm{P}_2$ is plotted as a function of temperature at three typical fields. A peaked feature, as marked by the down arrows at $T_{\rm N}$, is seen at three fields. Linear fits $1/T_1{\sim}T$ are shown for the 1.0~T, 2.0~T, and 4.0~T data below $T_{\rm N}$. 
}
\end{figure}

The low-field spin-lattice relaxation rate $1/^{31}T_1$ was measured on $\rm{P}_1$ and $\rm{P}_2$, as shown in Fig.~\ref{slrr}. Below 5~K, the P$_1$ and P$_2$ lines merge into a single peak due to inhomogeneous broadening (Fig.~\ref{spec}(a)). In this regime, $1/T_1$ was measured on the merged line, and the data are labeled as P$_1$ or P$_2$ according to which peak was being tracked during the cooldown sequence. With this labeling convention clarified, we now turn to the temperature evolution of $1/T_1$. Upon cooling below 50~K, $1/T_1$ decreases monotonically with temperature, and a kink in $1/T_1$ appears at 10~K, suggesting a phase transition or a crossover behavior. However, this temperature is about 5~K lower than $T_{x}$ as resolved by the peak feature in 
the $K_{\rm n}(T)$ [Fig.~\ref{spec}(b)]. Such differences indicate the onset of fluctuations with different energy scales; therefore, we attribute the phase to a short-range order antiferromagnetism (SROAFM) rather than the long-range order below $T_x$. Such an onset of SROAFM is typically seen in quasi-two-dimensional systems, where competing interactions cause a large entropy release at high temperatures without forming a long-range order\cite{Cui_DQCP_science}. 

Upon further cooling, a peak feature is resolved in $1/T_1$, as marked by $T_{\rm N}$ at about 1.8~K, signaling a second-order magnetic phase transition accompanied by critical spin fluctuations. Therefore, an ordered phase is suggested below $T_{\rm N}$. 
In the ordered state, $1/T_1$ follows a linear temperature dependence ($1/T_1{\propto}~T$) at 1~T, 2~T, and 4~T (see Fig.~\ref{slrr}), 
indicating that gapless low-energy excitation is a feature of the AFM1 phase away from the zero-field limit. This linear-$T$ behavior of $1/T_1$  is consistent with the reported supersolid Y phase in the single-layer, Ising triangular-lattice antiferromagnet Na$_2$BaCo(PO$_4$)$_2$~\cite{xuxiaoyu_prb}, although further confirmation would require single-crystal measurements.
With increasing magnetic field, the magnitude of $1/T_1$ at high temperatures is progressively suppressed, indicating the weakening of antiferromagnetic correlations. Meanwhile, the peak at $T_{\rm N}$ is progressively reduced and disappears near 4~T, indicating that the ordering temperature is driven to zero. This establishes a field-induced quantum phase transition at $H_{\rm C1} \approx 4$~T, as summarized in the phase diagram (Fig.~\ref{pd}). In addition, we note that $1/T_1{\propto}~T$ may not be a quantum critical behavior, as the $H_{\rm C1}$ is not precisely determined in this powered sample. The frustration factor is then estimated to be $\theta_{\rm CW}/T_{\rm N}~{\approx}$~20 at low fields, indicating strong magnetic frustration.


\section{NMR at intermediate fields}

\subsection{NMR spectra}

\begin{figure}[t]
\includegraphics[width=8.5cm]{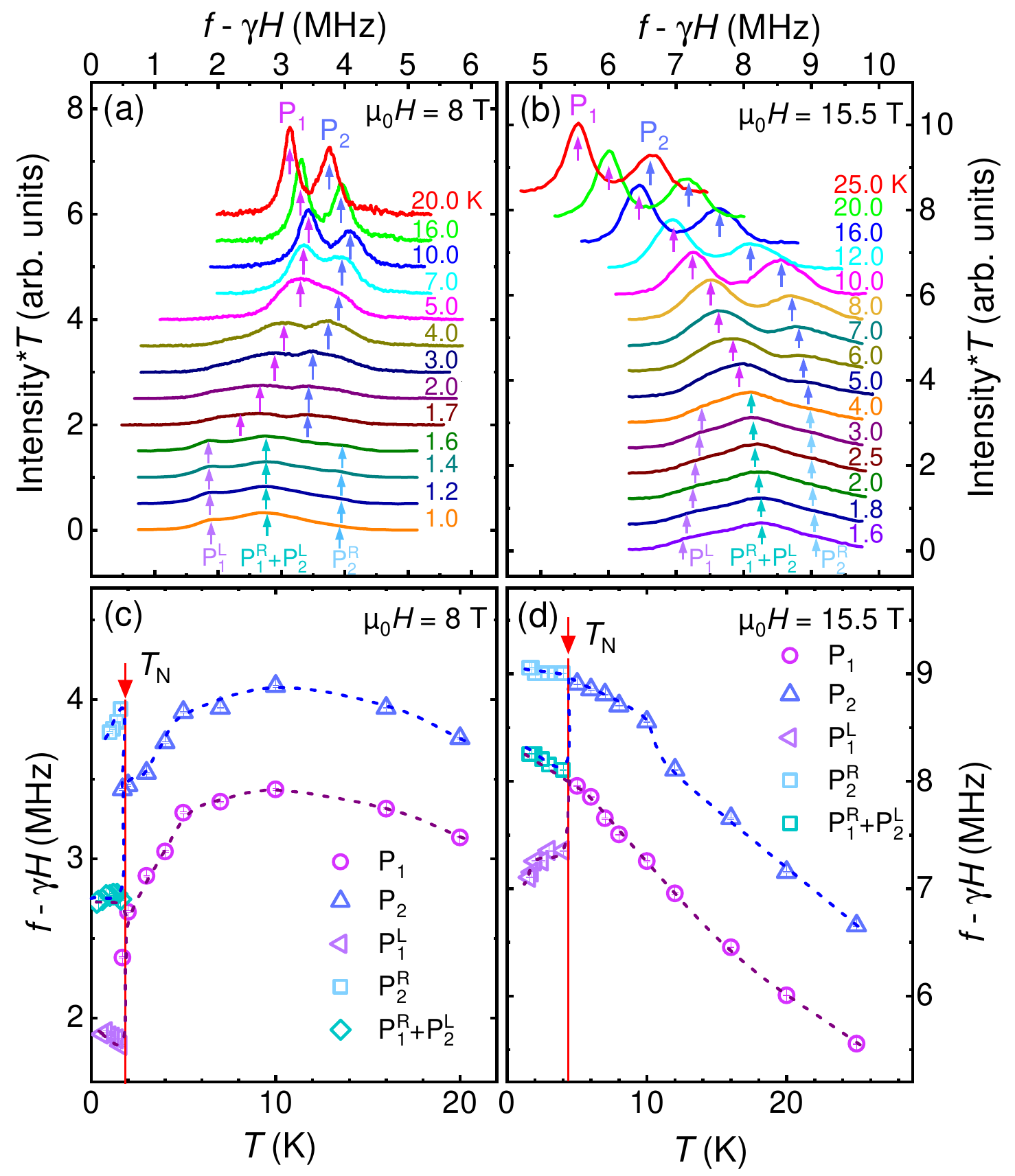}
\caption{\label{spec4}~\textbf{NMR spectra and line splitting at 8~T and 15.5~T.}
(a) The spectra at 8~T measured at typical temperatures. 
(b) The spectra at 15.5~T measured at typical temperatures. Vertical offsets are applied for clarity. 
At both fields, two peaks ($\rm{P}_1$ and $\rm{P}_2$) are resolved at high temperatures, while three peaks (${\rm P}^{\rm L}_1$, ${\rm P}^{\rm R}_1$+${\rm P}^{\rm L}_2$ and ${\rm P}^{\rm R}_2$) emerge at low temperatures.
(c)-(d) Temperature dependence of the resonance frequencies extracted from the spectra at 8~T and 15.5~T, respectively. The down arrows indicate the line-splitting temperature $T_{\rm N}$.
}
\end{figure}

\begin{figure}[h]
\includegraphics[width=7.5cm]{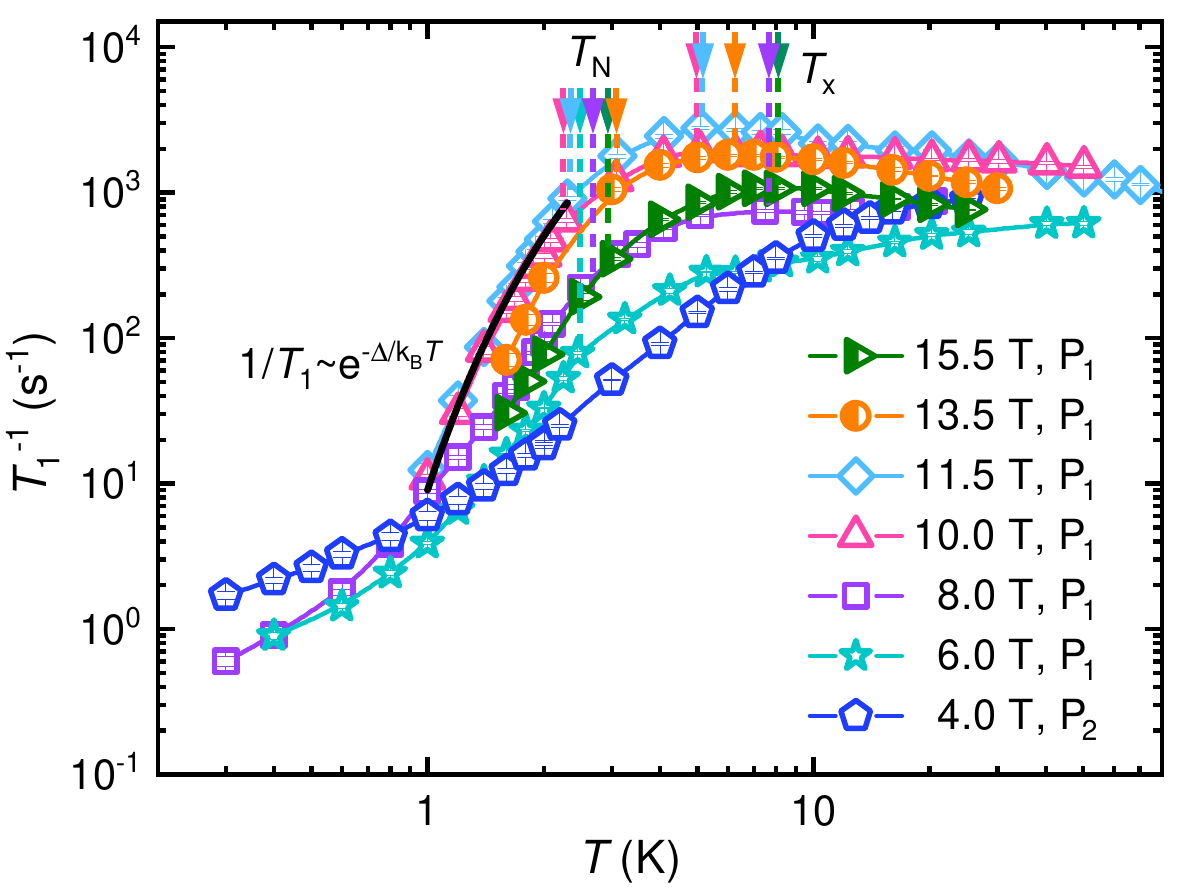}
\caption{\label{slrr2}~\textbf{Spin-lattice relaxation rate with field from 4~T to 15.5~T.}
$1/T_1$ as functions of temperatures under different fields measured on $\rm{P}_1$ and $\rm{P}_2$. $T_x$ marks the broad maximum in $1/T_1$, while $T_{\rm N}$ denotes the temperature below which $1/T_1$ decreases rapidly. The black solid line is a fit to the gap function $1/T_1{\sim}e^{-\Delta/k_BT}$ at temperatures below $T_{\rm N}$, where the spin gap is obtained as $\Delta$ = 7.73~K at 11.5~T. 
}
\end{figure}

Figure~\ref{spec4}(a) and (b) display the spectra measured at 8~T and 15.5~T, respectively, as functions of temperature. At temperatures above 6~K, two well-resolved peaks are observed, consistent with inequivalent $^{31}$P sites in the PM state. 
Below 6~K, the spectra broaden. Upon further cooling below 5~K, NMR line splitting is observed, where three peaks can be identified, clearly indicating the onset of a symmetry-breaking state. The peak frequencies at 8~T and 15.5~T are further extracted and displayed as functions of temperature in Fig.~\ref{spec4}(c)-(d). As $^{31}$P is a spin-1/2 nucleus which only couples to the magnetic moments, such further NMR line splits clearly demonstrate the onset of a second long-range ordered antiferromagnetic phase, which we term AFM2. The transition temperature, named as $T_{\rm N}$, determined by the line split temperature, is 2.0$\pm$0.2~K and 4.0$\pm$0.2~K, at 8~T and 15.5~T, respectively. $T_{\rm N}$ determined by the line split at each field is then plotted as a function of field in the phase diagram, as shown in Fig.~\ref{pd}. $T_{\rm N}$ first increases and then decreases with field. Such a domed feature of $T_{\rm N}$ is typical for field-induced magnetic ordering in the triangular lattice antiferromagnet~\cite{Cuiyi_PRM_2018}.

\subsection{Spin-lattice relaxation rate}

The $1/T_1$ measured under magnetic fields from 4~T to 15.5~T is shown in Fig.~\ref{slrr2}. Upon cooling, a broad maximum at $T_x$ is observed in most datasets. However, $T_x$ at each field is higher than the $T_{\rm N}$ determined by the NMR line splitting, indicating that this feature corresponds to a crossover into a short-range antiferromagnetic regime, similar to the decrease in the $K_{\rm n}$ data and the increase in the FWHM, rather than any long-range order.  

At lower temperatures, $1/T_1$ exhibits a sharp drop at $T_{\rm N}$, signaling the onset of long-range magnetic order. The extracted $T_{\rm N}$ values agree well with those determined from the NMR spectra and are summarized in the phase diagram (Fig.~\ref{pd}).

Below $T_{\rm N}$, $1/T_1$ follows a thermal activation behavior, $1/T_1$ $\sim e^{-\Delta/k_BT}$, indicating the opening of a spin gap. At 11.5~T, the gap is obtained as $\Delta~{\approx}$~7.73~K. 
The gap remains nearly field-independent up to $\sim$18~T, and then increases rapidly at higher fields [see Fig.~\ref{gap}(b)]. Such gapped behavior may be caused by the field: with applied field, the most probable orientation of the sample is with the field along the $ab$-plane, which breaks the rotational symmetry. 

The field dependence of $1/T_1$ at a fixed temperature below $T_x$, for example at 3~K, shows a nonmonotonic behavior: $1/T_1$ increases with field up to 11.5~T and decreases at higher fields. This indicates that low-energy spin fluctuations are maximized near 11.5~T, consistent with a dome-shaped evolution of $T_{\rm N}(H)$, although not well established by our data.

\section{High-field NMR}

\begin{figure}[t]
\includegraphics[width=8.5cm]{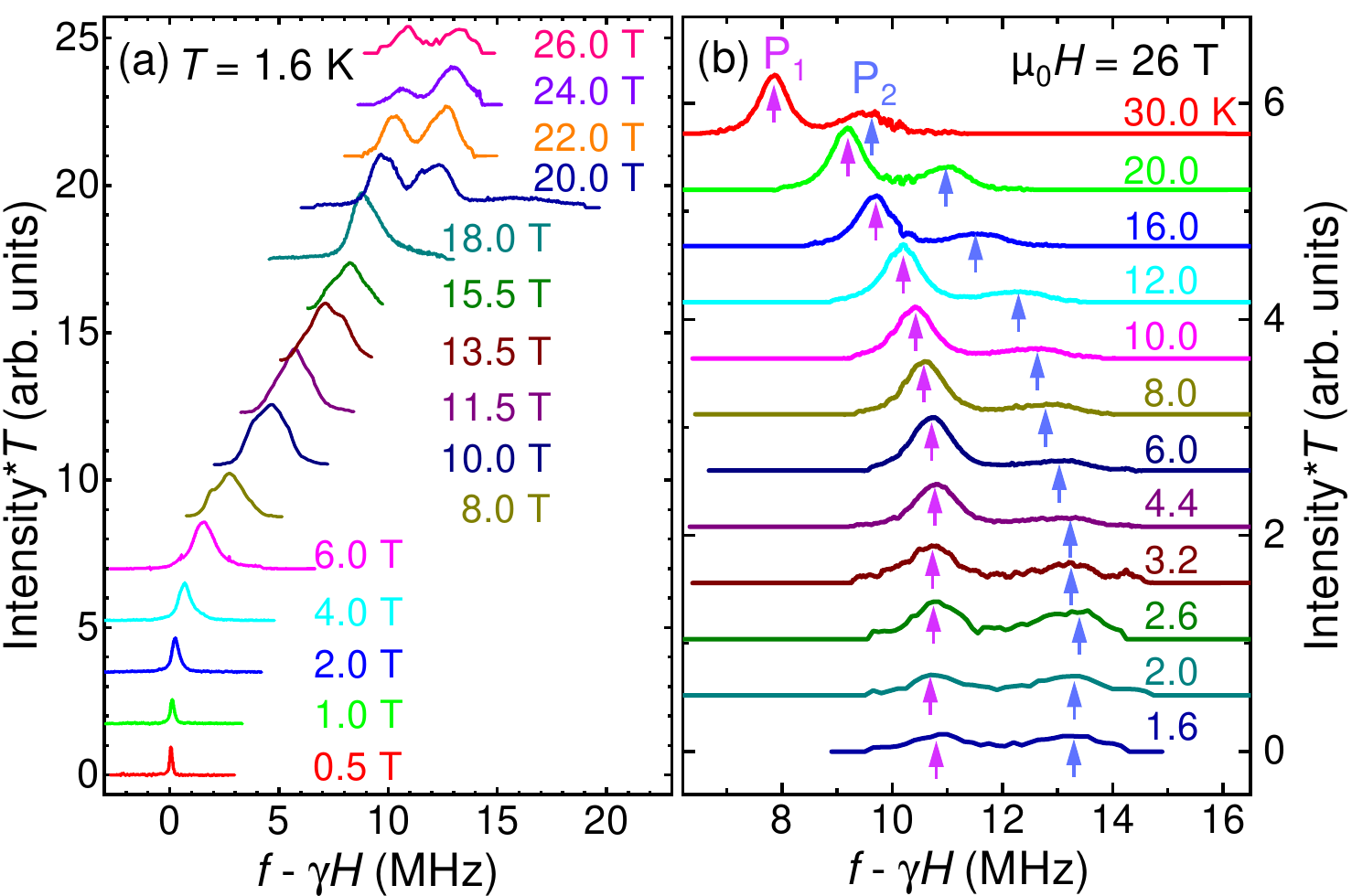}
\caption{\label{spec3}~\textbf{NMR spectra with field up to 26~T.}
(a) The lowest temperature spectra. 
(b) Spectra measured at 26~T at typical temperatures.  Vertical offsets are applied for clarity. 
}
\end{figure}



\subsection{NMR spectra}

To determine the phase boundary of the ordered phase, high-field NMR is conducted with field up to 26~T. The spectra collected at a constant temperature of 1.6~K are presented with field from 0.5~T to 26~T in Fig.~\ref{spec3}(a). 
At low fields (0.5–4~T), the spectra exhibit a single peak, corresponding to the low-field antiferromagnetic phase (AFM1) in Fig.~\ref{pd}. In the intermediate field range (4–18~T), the spectra split into three peaks, indicating the emergence of  distinct ordered phases. 


For fields above 20~T, the NMR line splits into two peaks; however, this is indeed caused by inequivalent $^{31}$P sites rather than magnetic ordering, as verified in Fig.~\ref{spec3}(a). At our highest field of 26~T, the spectra were collected with temperature up to 30~K. The double-peak structure persists up to 30~K, indicating that it arises from crystallographically inequivalent P sites rather than from magnetic ordering.

\subsection{Spin-lattice relaxation rate and spin gaps}

\begin{figure}[t]
\includegraphics[width=8.5cm]{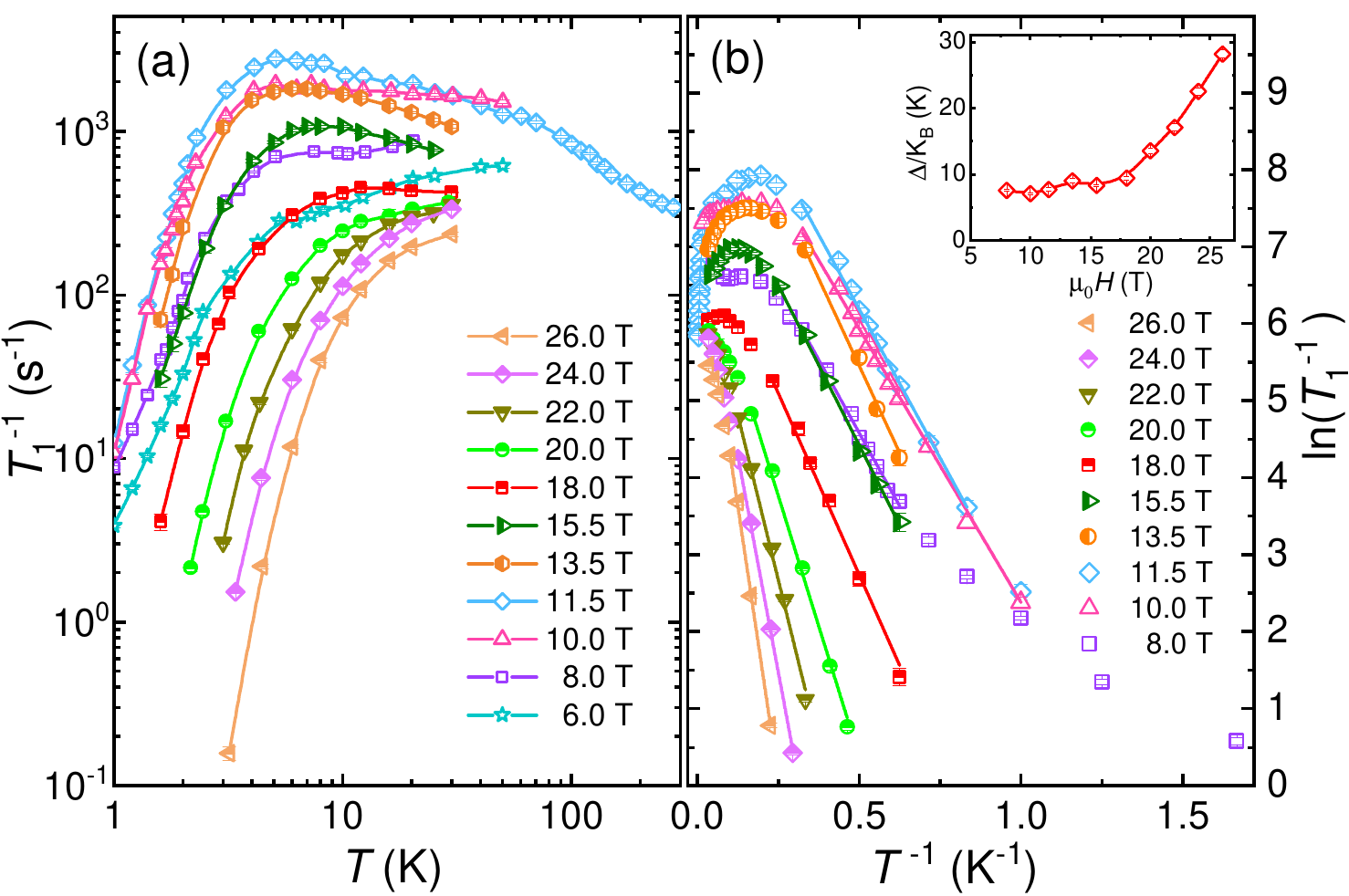}
\caption{\label{gap}~\textbf{High-field spin-lattice relaxation rate.}
(a) $1/T_1$ as functions of temperatures with fields from 6~T to 26~T.
(b) Semilog-plot of $1/T_1$ as functions of $1/T$. The solid lines are linear fits to the data; the slope at each field determines the gap value. Inset: Gap $\Delta$, obtained from the fits in the main plot, plotted as a function of field. 
}
\end{figure}

To further identify the phase boundaries, $1/T_1$ was measured over a wide field range from 6~T to 26~T, as shown in Fig.~\ref{gap}(a). The overall temperature dependence of $1/T_1$ in the high-field regime is similar to that at intermediate fields, exhibiting a pronounced suppression at low temperatures.

To quantify this, we analyze the data using a thermally activated form, $1/T_1 = a e^{-\Delta/k_BT}$, where $\Delta$ is the spin gap. Accordingly, $\ln(1/T_1)$ is plotted as a function of $1/T$ in Fig.~\ref{gap}(b), where a linear dependence is observed within a limited temperature window. The spin gap $\Delta$ is extracted from the slope of the linear fits.

The extracted gap values are summarized in the inset of Fig.~\ref{gap}(b) as a function of magnetic field. The gap remains nearly constant with field up to 18~T and increases for fields above 18~T, indicating an enhancement of the spin gap in the high-field FP phase. Therefore, a second transition field is suggested at $H_{\rm C2}~{\approx}$~18~T.
Unfortunately, we did not resolve the gapless phase by the $1/T_1$, and therefore the transition fields among the 1/3-plateau, the 2/3-plateau, and the fully polarized phases are not determined. This may arise from the polycrystalline nature of the material, making it difficult to determine the phase boundaries precisely.

\section{Phase diagram and discussions}

\begin{figure}[t]
\includegraphics[width=8.5cm]{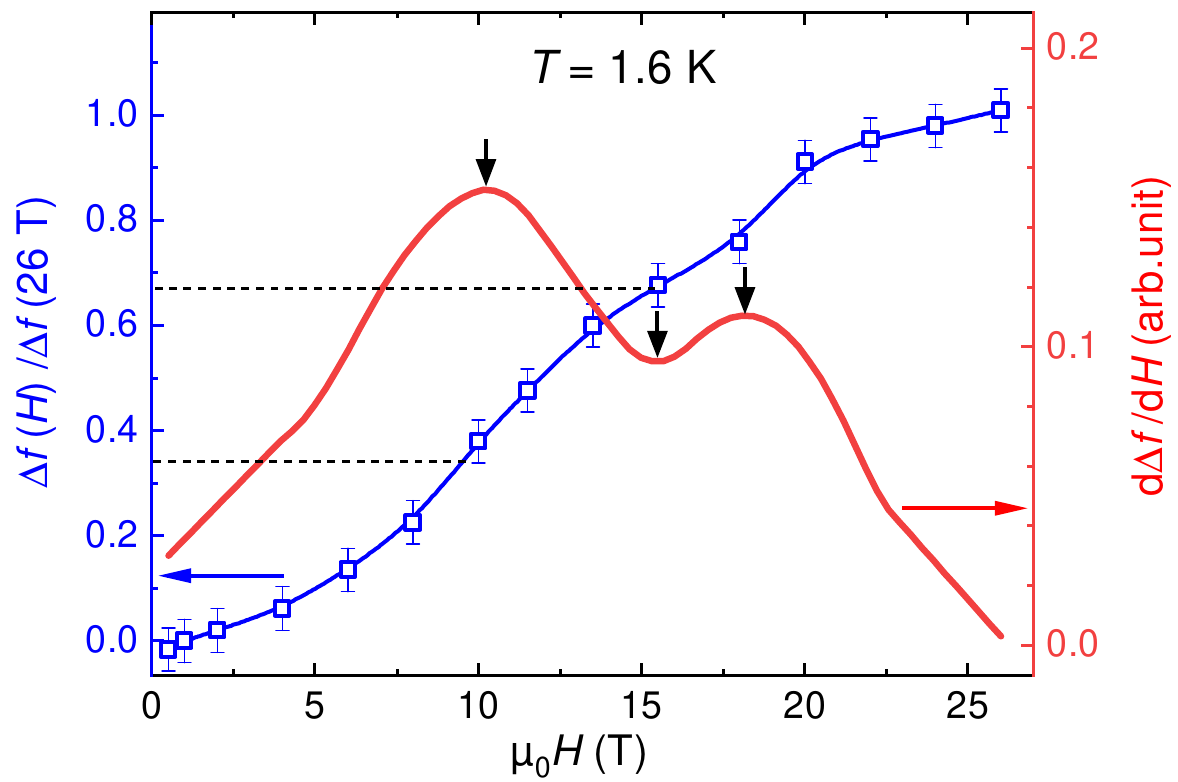}
\caption{\label{MP}~\textbf{NMR frequency shift.}
Relative NMR frequency shift ${\Delta}f$ = $f-{\gamma}H$ at 1.6~K is shown on the left axis as a function of fields. The frequency $f$ is the frequency deduced by the average frequency of the NMR spectra. The dashed lines present the 1/3 and 2/3 of full saturation of frequency, which intercept with the ${\Delta}f$ line at 10~T and 15.5~T respectively.   
The differential frequency $d{\Delta}f/dH$ as a function of field is shown on the right axis. The arrows mark the peak and the dip positions on the $d{\Delta}f/dH$ curve, located at 10~T, 15.5~T, and 18~T,  which correspond to the phases with 1/3 and 2/3 of the fully saturated moments, and the transition field to the paramagnetic phase, respectively.
}
\end{figure}

\begin{figure}[t]
\includegraphics[width=8.5cm]{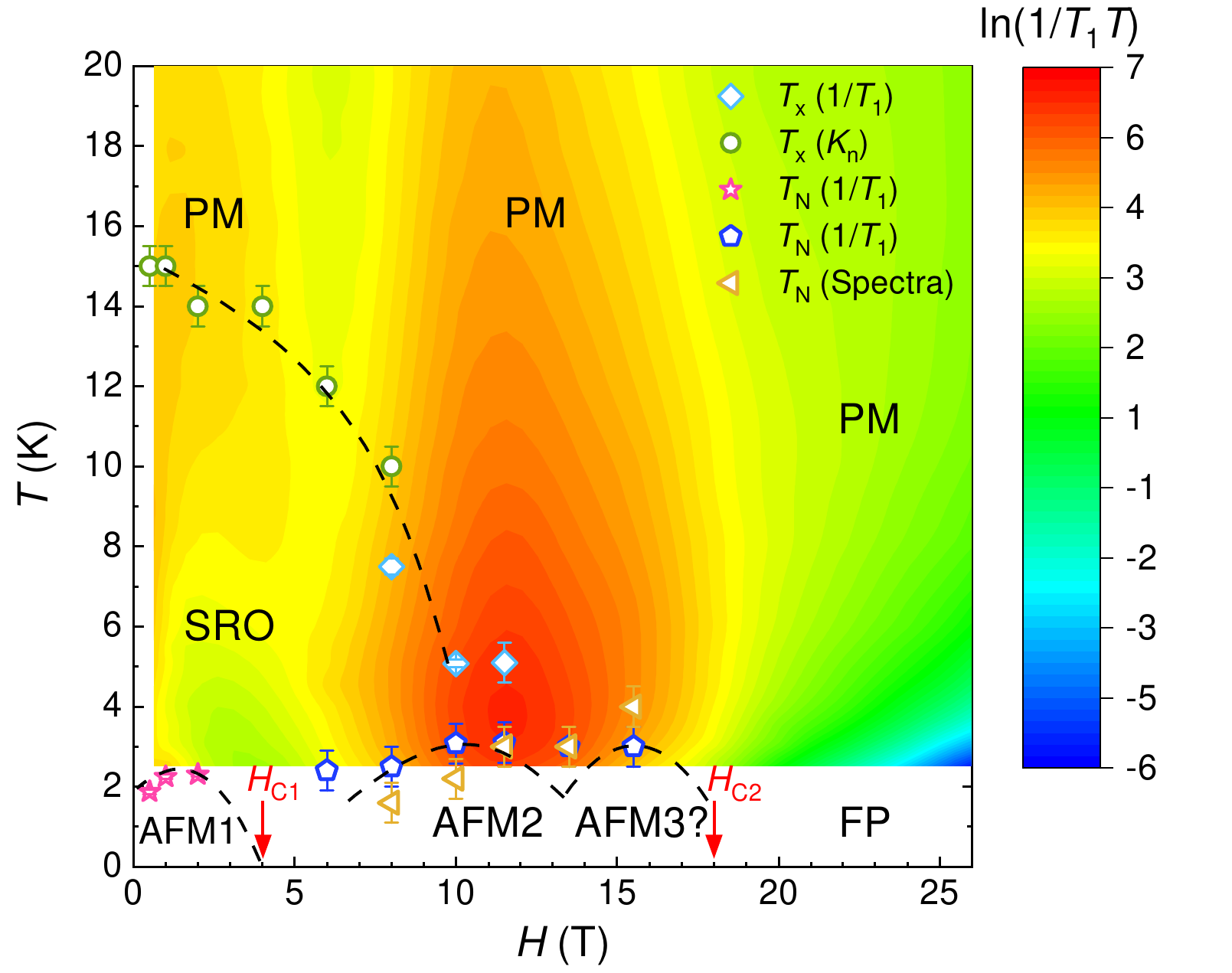}
\caption{\label{pd}~\textbf{Magnetic phase diagram overlaid on the contour map of $\ln(1/T_1T)$.}
The phase diagram and phase boundaries are determined by the NMR spectra and the $1/T_1$ as specified. Here the AFM1 and AFM2 phases are proposed to correspond to the supersolid and the 1/3-plateau phases, respectively. The 2/3-plateau phase may exist as labeled by AFM3. A short-range-ordered regime is seen at temperatures below $T_x$. The fully polarized (FP) phase is established at fields above 20~T. 
}
\end{figure}

As the NMR spectra directly reflect the local spin polarization, the field dependence of the resonance frequency provides a reliable measure of the magnetization process.
The frequency shift, defined as ${\Delta}f$ = $f-{\gamma}H$, is then calculated at each field at a fixed temperature 1.6~K, as shown in Fig.~\ref{MP}. Here $f$ is the average frequency of the NMR spectra. Note that ${\Delta}f(H){{\propto}M(H)}$, where $M(H)$ is the bulk magnetization of the material. ${\Delta}f$ reaches a constant value with field from 20~T to 26~T, indicating the fully polarized (FP) state.
The derivative $d(\Delta f)/dH$ shown in Fig.~\ref{MP} was computed by first fitting the discrete $\Delta f$ data with a smooth interpolation curve, then taking the analytical derivative. This smooth derivative curve is presented only as a guide to the eye for identifying the approximate positions of slope changes; the indicated peak and dip fields ($\sim$10~T, $\sim$15.5~T, $\sim$18~T) should be taken as qualitative, and higher-resolution measurements would be needed to determine critical fields precisely.

The ${\Delta}f(H)$  curve exhibits clear non-linear behavior with increasing field, and more importantly, shows two plateau-like features at intermediate fields.
Quantitatively, the magnetization reaches approximately 1/3 $M_{\rm S}$ around 10~T, where a pronounced peak in $dM/dH$ is also observed, indicating a field-induced stabilization of a fractional magnetization state. Upon further increasing the field, ${\Delta}f(H)$  shows another flat region around 15.5~T, corresponding to approximately 2/3 $M_{\rm S}$. These two characteristic features strongly suggest the emergence of the 1/3 and 2/3 magnetization plateau phases.
Therefore, the 1/3 and the 2/3-plateau phases may be realized in the system, which needs to be verified by high-field magnetization measurements. 

A contour plot of $\ln(1/T_1T)$ over the field range 0.5–26~T is shown in Fig.~\ref{pd}. Based on the NMR determination of phase transitions and crossovers, a comprehensive $(H,T)$ phase diagram is constructed and overlaid on the color map. Several distinct magnetic phases are identified as a function of field, including the low-field AFM1 phase, the field-induced AFM2 phase, the fully polarized (FP) phase, and a possible intermediate AFM3 phase. 
Their phase transition temperatures are also determined by different methods as specified. In addition, a high-temperature short-range-ordered phase is also revealed by the onset of 
the increase of the NMR linewidth. Notably, the phase diagram is fully consistent with the evolution of low-energy spin fluctuations: the $1/T_1T$ exhibits the strongest high-temperature spin fluctuations at about 11.5~T, consistent with the largest $T_{\rm N}$ revealed by the NMR line splitting.

The low-field NMR supports a gapless behavior for the AFM1, which may be consistent with the onset of supersolid phases~\cite{Sakai_PRB_2011}. On the other hand, the gapped behavior in the high-field ordered phase could be caused either by the field-induced in-plane symmetry breaking or the proximity to the magnetization plateaus~\cite{Xu_CM_2024}. We note that ${H_{\rm C1}~{\approx}}$~4~T marks the AFM1--AFM2 boundary, while ${H_{\rm C2}~{\approx}}$~18~T corresponds to the AFM3--FP transition (Fig.~\ref{pd}); the 1/3 and 2/3 plateau features at 10~T and 15.5~T lie within the AFM2 phase. However, note that the 2/3-plateau phase may be caused by strong bilayer coupling, as such a plateau phase is rarely seen in single-layered triangular lattice antiferromagnet~\cite{li_SR_2015,Momoi_PRB_1999}. Nevertheless, our results suggest that the dimerization inside the bilayers is not strong enough to induce a spin-gapped dimer phase. Recently, a type of bilayered triangular lattice antiferromagnet, Rb$_2$Co$_2$(SeO$_3$)$_3$, has been discovered~\cite{Xu_CM_2024}, which exhibits strong bilayer dimer coupling. This compound shows field-induced 1/3, 1/2, 2/3 and 5/6 plateau phases when the field is applied along the crystalline $c$ axis~\cite{xuxiaoyu_prb,Fu_PRB_2025}. Future studies on the single crystals of the current compound are also requested to resolve the magnetic anisotropy and novel magnetic plateau phases. We also emphasize that powder NMR alone cannot determine the precise spin configurations; complementary thermodynamic probes such as specific heat and torque magnetometry on single crystals will be valuable for this purpose.  

\section{Summary}

In summary, we performed a $^{31}$P NMR study on the bilayer triangular lattice antiferromagnet K$_2$Co$_2$(HPO$_3$)$_2$(C$_2$O$_4$)(H$_2$O)$_2$, revealing several magnetic phases and phase transitions with increasing field. At fields below 4~T, a gapless ordered ground state is suggested, which is a candidate spin supersolid phase. Upon increasing the magnetic field, the system undergoes a quantum phase transition at a critical field ${H_{\rm C1}~{\approx}}$~4~T. Above 4~T, field-induced magnetically ordered phases AFM2, emerge as evidenced by the clear NMR line splitting. A spin gap is also resolved in the ordered phases with fields from 5~T to 18~T, before the system becomes fully polarized. The differential magnetization, as revealed by the relative NMR frequency shift, exhibits the peak and the dip features at 1/3 and 2/3 of full saturation moments. Our results establish K$_2$Co$_2$(HPO$_3$)$_2$(C$_2$O$_4$)(H$_2$O)$_2$ as a unique bilayer platform for studying field-induced quantum phase transitions in frustrated magnets. The transition from a low-field gapless ordered state to a field-induced ordered phase highlights the delicate balance between frustration, quantum fluctuations, and magnetic field in determining the ground state of low-dimensional quantum magnets.

{\bf Acknowledgments.} 
This work was supported by the National Key Research and Development Program of China (Grant No.~2023YFA1406500, No.~2025YFA1412100, No.~2022YFA1403402 and No.~2024YFA1409200),
National Natural Science Foundation of China (Grants No.~12374156, No.~12134020 and No.~22422111), the Scientific Research Innovation Capability Support Project for Young Faculty (Grant No.~ZYGXQNJSKYCXNLZCXM-M26), and the Strategic Priority Research Program of Chinese Academy of Sciences (Grant No. XDB1270100). A portion of this work was carried out at the Synergetic Extreme Condition User Facility (SECUF, https://cstr.cn/31123.02.SECUF).

\bibliographystyle{iopart-num.bst}
\bibliography{ref}

\end{document}